\documentstyle[12pt]{article}

\newcommand \be {\begin{equation}}
\newcommand \ee {\end{equation}}
\include{dspace12}
\include{epsf}
\include{mpla1.sty}
\begin{document}
\begin{center}
{\bf{\large Conformal transformations near Naked Singularities - I}}
\end{center}

\begin{center}
{Sukratu  Barve}\footnote{sukratu@imsc.ernet.in}\\
  The Institute  of Mathematical Sciences\\
   C.I.T. Campus\\
   Taramani\\
   Chennai 600 113\\
   INDIA\\ 
\end{center}

\abstract
          {We show  that  the behaviour of  the outgoing radial null
 geodesic congruence on the boundary of the trapped region (suitably defined 
 as a four dimensional region) is  related to the  property
 of nakedness in spherical dust collapse. The argument 
 involves a conformal 
 transformation which justifies the  difference in 
 the Penrose diagrams in the naked and covered dust collapse scenarios.}
\newpage
\section{Introduction}

        Consider a cloud of matter (regular initial Cauchy data) collapsing indefinitely under its own gravity. A singularity eventually develops in the spacetime and it is indicated by the divergence of the Kretschmann scalar. In the advanced stages of collapse trapped regions are formed \cite{schoen}, \cite{penrose}
 and there exists a null ray which marginally escapes to infinity (event horizon).
 It is not clear whether the singular boundary is entirely surrounded by the trapped region. In other words, it is not known if a portion of the singular 
boundary is exposed in the untrapped region and non-spacelike geodesics can 
emanate from it (naked singularities).
In fact, exact solutions to Einstein's field equations with certain kind of 
source 
terms are known to exhibit both naked and covered singularities depending upon
 the sort of regular initial data chosen (See \cite{rev1} for details).

 The complexity of the general problem lies in the fact that the Cauchy initial
 value problem for the Einstein field equations with sources is less
 tractable. The systematics available about the initial value problem is far 
too
 small for any implication for questions like formation of naked singularities.
  For instance,
 even the well-posedness of the problem is not self evident and has to
 be proved independently for different types of sources \cite{waldgr}, \cite{haw}, \cite{gara}. It would be indeed difficult, for example, to find or even expect
 a conserved or monotonically behaving function of the Cauchy surface
 with respect to its evolution, which could be expected to provide
 insight into the process of creation of naked singularities.

   As a result of this difficulty, a large number of investigations that have been carried out, have been 
 concerning  certain exact solutions or numerical simulations (\cite{rev1} and references cited in them). Issues like
  strength of the singularities, genericity, behaviour
 with respect to change of source etc. have been studied in examples
 like dust, null dust, perfect and imperfect fluids and scalar fields.
 However, they do not suggest any 
 typical geometrical feature which could be expected to arise before
 a naked singularity forms ( In case of a singularity the singularity theorems
  make use of a typical geometrical feature viz. trapped surfaces to prove its existence).  
 The lack of indication of existence of such a feature
 is evident in the fact that one is forced to check for the existence of naked
 singularities
 in a direct manner whenever required. To be precise, one simply checks if
non space-like geodesics emerge from the singular boundary using differential
 geometry and the form of the metric in the example.  

This paper is a first step towards an indirect criterion. Preferably, the criterion should be applicable away from the singular boundary. That is a non-local
 problem and given the difficulties of Cauchy evolution, there is no indication available for what the criterion could be.\\ It would be perhaps appropriate in such a situation to examine regions near the singularity. There are also parts of the 
 singularity from
 which geodesics cannot escape. Some criterion applicable to such a portion
 would also be significant. It would indicate that the information about
 the exposure of a part of the boundary is contained elsewhere on the boundary.

 This work is the first part in which the example of spherical dust collapse is
 considered. This example is well studied and the initial data which leads to
 naked singularities is known.
 
\section{Spherical Dust Collapse}
We illustrate the main features of spherical dust collapse in radial co-ordinates (figures 1 and 2) and in Penrose diagrams (figures 3 and 4).

\begin{figure}[!h]
\parbox[b]{10.99cm}
{
\epsfxsize=6.95cm
\epsfbox{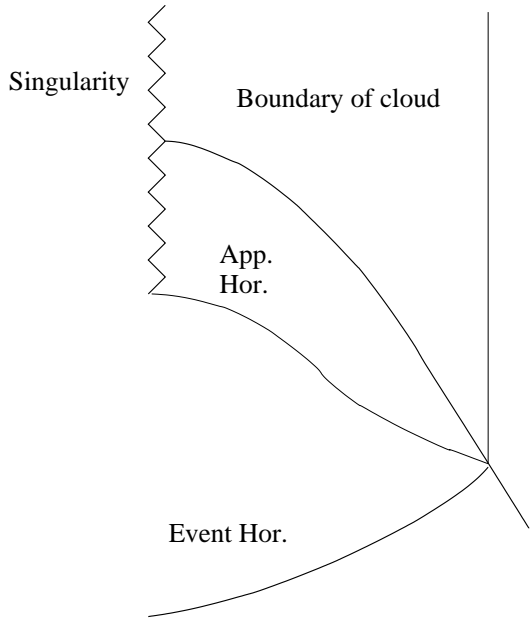}
}
\caption{Collapse of spherical dust leading to a covered singularity}
\end{figure}

\begin{figure}[!h]
\parbox[b]{10.99cm}
{
\epsfxsize=6.95cm
\epsfbox{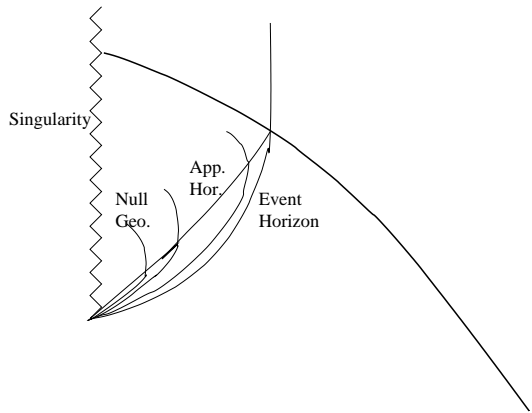}
}

\caption{Collapse of spherical dust leading to a naked singularity}
\end{figure}

\begin{figure}[!h]
\parbox[b]{10.99cm}
{
\epsfxsize=6.95cm
\epsfbox{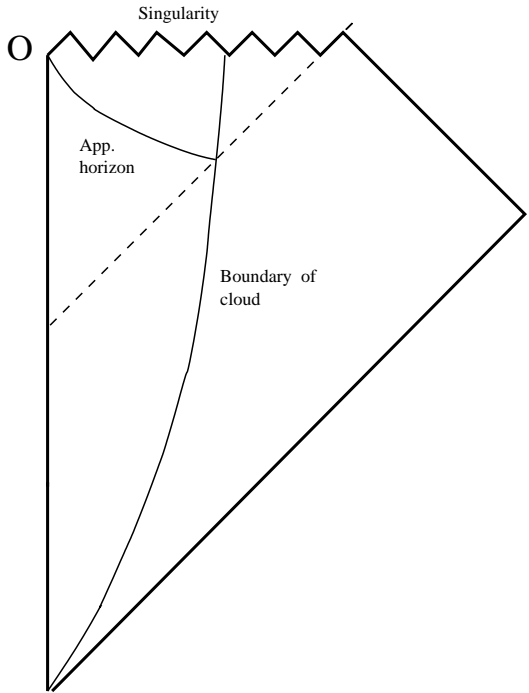}
}
\caption{Penrose Carter diagram for collapse of spherical dust leading to a 
covered singularity}
\end{figure}

\begin{figure}[!h]
\parbox[b]{10.99cm}
{
\epsfxsize=6.95cm
\epsfbox{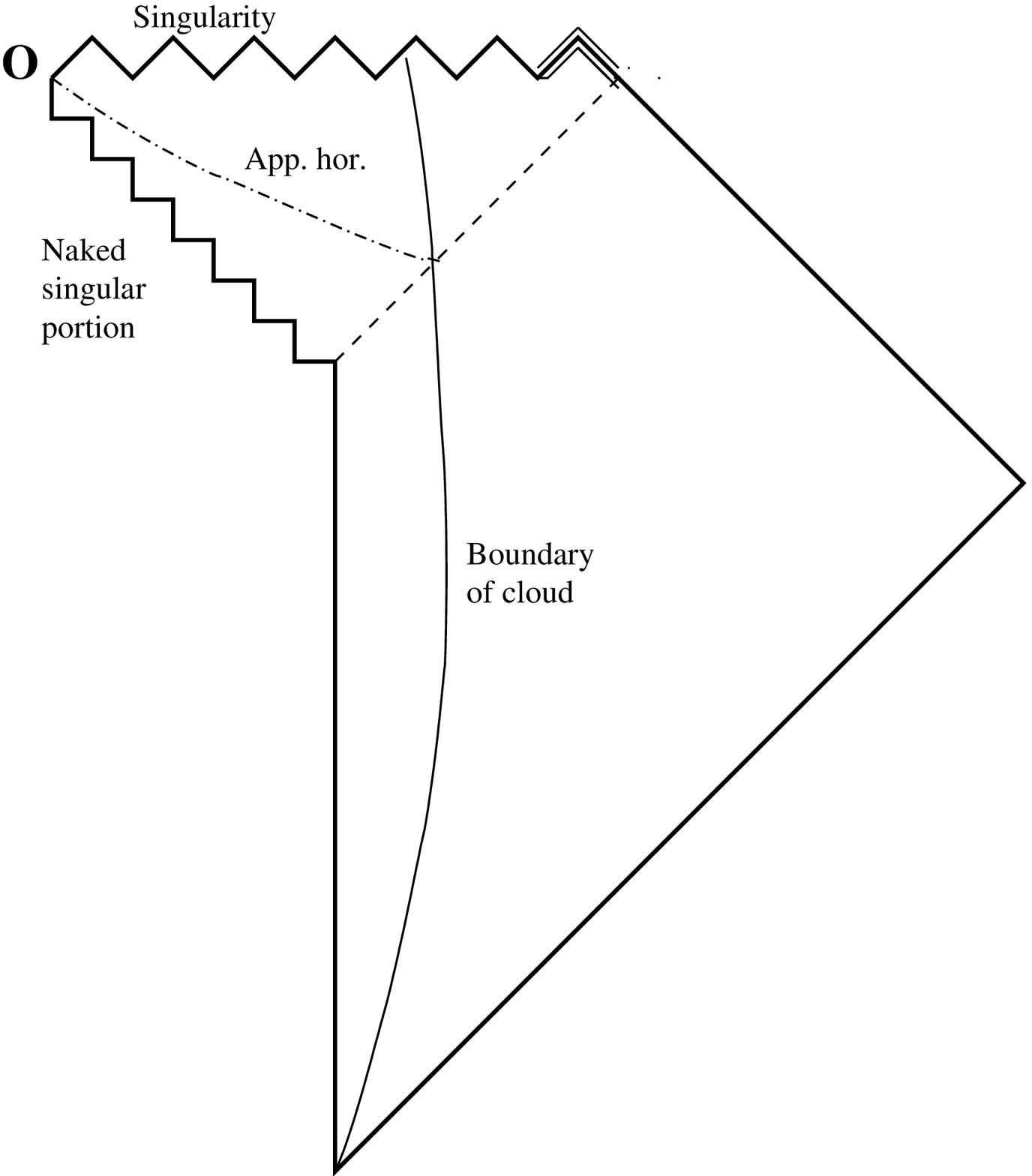}
}
\caption{Penrose-Carter diagram for collapse of spherical dust leading to a 
(locally) naked singularity}
\end{figure}

Point O is defined as the point where the apparent horizon\footnote{
 In this work we define a trapped region as the union of all possible closed
trapped
 surfaces (See \cite{haw} or \cite{waldgr} for the definition of a closed 
trapped 
surface). This is in general a four dimensional region. The boundary of this 
region within spacetime is referred to as `apparent horizon' in this work. 
 However, according to the usual definition (See \cite{haw} and \cite{waldgr}) , a section of this boundary by a Cauchy surface would be called an apparent
 horizon. We make this change for lack of established terminology for the 
ideas we use (which are independent of any Cauchy surface). See also figure
 60 of \cite{haw} and \cite{jhingan}.} meets
 the singularity.
In figure 3, there is no portion of the boundary beyond O that is exposed. In figure 4, however, there is a null portion.
 In this paper, we examine if O would yield the information about the
 existence of an exposed portion. 
  O is a covered point of the singular boundary (with or
 without a naked portion).  Mathematically, one works with points on the apparent
 horizon in the approach to the concerned point O. This could be
 looked upon as a property of the marginal trapped regions which constitute
 the apparent horizon. One is therefore working with
 outgoing  null congruences with zero expansion. If a singularity meets
 such a region, the point in the strict sense will not have any non-spacelike geodesics emerging
 from itself and will therefore be covered. 

 We proceed as follows. Consider a congruence of outgoing null geodesics.
 Let us parametrize each geodesic using an affine
 parameter. We investigate the tangent vectors {\bf$\xi$} to geodesics in the congruence
 at the apparent horizon ($\xi^{\mu}=dx^{\mu}/ds$ where $s$ is an affine parameter along the null geodesic of the congruence). 

O is a part of the
 singularity which is marginally trapped (expansion of any outgoing congruence  will be zero here)  and so there will be no outgoing geodesic emerging from O. Any metric which confirms to that is referred to in this paper as  having the 
 `correct' causal structure at the boundary.

In the limit of approach to O,
 the vector {\bf$\xi$} is expected to vanish if the metric used depicts the appropriate
 causal structure at O. O is a part of the
 singularity which is marginally trapped (expansion of any outgoing congruence  will be zero here) there will be no outgoing geodesic emerging from O. 
 Any metric which confirms to that is referred to in this paper as  having the 
 `correct' causal structure at the boundary.    
We examine this in the case of spherical dust collapse where the metric is
 evolved with regular initital Cauchy data. It is found that
 the condition is met at O for the collapse leading to covered singularities.
 However, the tangent vector stays non-zero in the limit for the naked case.
  This is remedied by performing a conformal transformation which
 diverges at O. The appropriate causal structure is then restored.
  Thus, in case of spherical dust collapse, this behaviour
 of the tangent vector is related to the property of nakedness of the 
singularity. Could this be true in general? It would mean that a diverging
 conformal tansformation is necessitated at the singular boundary in
 the naked case as against the covered case. Some theorems in the case of 
general collapse 
 are provable and work is in progress. 

{\it Thus, it is suggested that in a general collapse, the tangent vector field of the congruence on the apparent horizon (precisely, its limiting behaviour at the trapped part of
 the singularity) contains the information about whether the singular boundary
 has a naked portion.} 

  The plan of the rest of the paper is as follows. We wish to investigate the
 tangent 
vector on the apparent horizon. We choose the radial null geodesic congruence
 in spherical symmetry. The general expression
 for tangent vector is not available in a closed analytic form. However,
 one can calculate it in a simpler special case of dust, the self similar model, and
 subsequently show how to generalize the results.
 The next section describes the self similar dust model,  where we discuss the
 tangent vector
 and demonstrate the connection with nakedness. In the next section  we show that the results
 can be extended  to the general dust case. We then  turn to the mentioned 
 conformal  transformation in the next section. It leads to the Penrose diagram
 for the naked case from the usual metric of Tolman and Bondi in spherical 
co-ordinates. We show that the conformal transformation should diverge at the 
singular boundary. This results from the fact that the tangent vector ought to 
 vanish at O in the correct causal metric.

\section{Self Similar Dust Model}
   
   The collapse of a spherical cloud of pressureless fluid
 is given by the following metric \cite{rootspap}.

 \be
 \label{gendust}ds^{2}=dt^{2}-\frac{R'^{2}}{1+f(r)}dr^{2}-R^{2}d\Omega^{2}
\ee

 where 

$t$ and $r$ are the co-moving time and radial co-ordinates respectively.
$R(t,r)$ is called the `area radius' and a closed expression for
 this quantity which results in the dust case has enabled substantial
 progress in understanding the model.

Two free functions arise viz. $F(r) $ , called the mass function since it is
 the total mass to the interior of a shell of radius $r$ and the total
 energy function $f(r)$ which is called so because of the constraint below
which resembles a relation between kinetic and gravitational potential
 energies of a shell.
\be
\dot{R}^{2}=\frac{F}{R}+f
\ee
The source energy momentum tensor is $diag[\rho(t,r),0,0,0]$.

The solution for $R$ mentioned above is
\be
t-t_{0}(r) =-\frac{R^{3/2}}{\sqrt{F}} {\large G}\left(\frac{-Rf}{F}\right)
\ee
   
where a singularity boundary is formed at $t=t_{0}(r)$. The function $G$
 is defined as follows.

 $$
   G(y)= \cases { \frac{{\rm arcsin}\sqrt{y}}{y^{3/2}} -
                     \frac{\sqrt{1-y}}{ y}, &  $1\geq y> 0$,\cr
                     {2\over 3}, & $y=0$,\cr
                     {-{\rm arcsinh}\sqrt{-y}\over (-y)^{3/2}} -
                     {\sqrt{1-y}\over y}, & $0> y\geq -\infty$.\cr}           
 $$

The central
 shell focussing singularity, which is the limit as $r \rightarrow 0$
 along this locus is of interest and turns out to be naked for some initial
 data. 

 The self-similar model is the one in which $F(r)=\lambda r $ where 
 $\lambda $ is a constant (which decides if the central singularity will
 be naked or not) and $f(r)=0$.
 \footnote{ The model is referred to as self similar since there exists a 
 homothetic Killing field $t\frac{\partial}{\partial t}+ r\frac{\partial}{\partial r} $}

 We choose the scaling $t_{0}(r)=r$. A self similar co-ordinate $z=t/r$ is
 introduced. We note the expressions for $R$ and $R'$ which will be useful
 in the subsequent analysis

 \be
 R= r \lambda^{-2/3}\left(3/2\left(z-1\right)\right)^{2/3}
\ee
\be
 \label{R'eqn}R'= \left(\frac{2\lambda/3}{z-1}\right)^{1/3}\left(\frac{z-3}{2}\right)
\ee

 We cast the metric into double null co-ordinates. It is not difficult 
 to show that
\be
ds^{2}=r^{2}\left(z^{2}-R'^{2}\right)~~du~~dv
\ee
where
\be
\label{du} du=\frac{dr}{r}+\frac{dz}{z-R'(z)}
\ee
\be
dv=\frac{dr}{r}+\frac{dz}{z+R'(z)}
\ee

The double null form ($ds^{2}=C^{2}(u,v)dudv$) turns out to be useful when affine  parameters
 along null geodesics are to be calculated. For instance, along an 
 outgoing radial null geodesic ($du=0$), the affine parameter is
 $\int_{u=constant}C^{2}dv$ up to a multiplicative and an additive constant.

Now let us turn to calculating the tangent vector to the outgoing null
 radial geodesic congruence, which is our primary interest.

Assume the vector to be of the form

\be
\label{xieqn}  {\bf \xi} = (Q(t,r), Q(t,r)\sqrt{1+f}/R',0,0)
\ee

where $Q$ is obtained from the geodesic equation which {\bf $\xi$} has to
 satisfy. That constraint turns out to be

\be
\label{Qeqn} Q\dot{Q} + QQ' \sqrt{1+f}/R' + Q^{2}\dot{R}'/2R' = 0
\ee 

We have provided the expressions for the most general dust case here.
 One may read off the expressions for the self similar case by setting
 $f$ to zero and using equation (\ref{R'eqn}) for $R'$.

The equation above takes the form

\be
\label{1/Q} 1/Q=\int_{u=constant}\frac{\dot{R'}}{2R'}dk ~~+~~A(u)
\ee
 where 
 $k$ is an affine parameter along outgoing radial null geodesics 
 and $u$ is the retarded null co-ordinate. $A$ is an arbitrary function
 of $u$ resulting because of the partial integration. 

 As indicated earlier,
\be
dk = r^{2}\left(z^{2}-R'^{2}\right)dv
\ee 
keeping $u$ fixed. 

Using this in equation (\ref{1/Q}) we obtain

\be
1/Q= \int_{u=constant} \frac{r}{3}\frac{\left(z-3\right)^{2}}{z^{2}-1}\left[\frac{2\lambda/3}{z-1}\right]^{1/3}~~dz
\ee

 or using the fact that $du=0$ from equation (\ref{du})~~\footnote{ The expression
 for $r$ thus obtained in terms of $z$ has a positive multiplicative constant.
  Whether $r$ vanishes or not( and later whether
 $1/Q$ vanishes or is non-zero) is of importance in the calculations. So setting this constant to $1$, which is done here, will not affect the result.} 

\be
\label{1/Qint}1/Q= \int_{u=constant} \frac{1}{3}\frac{\left(z-3\right)^{2}}{z^{2}-1}\left[\frac{2\lambda/3}{z-1}\right]^{1/3} e^{\left[-\int\frac {db}{b-R'(b)}\right]_{z}}~~dz
\ee

The integral over $z$ is to be evaluated from $r=0$ to the apparent horizon, where we shall be interested in evaluating the tangent vector.
 The latter can be shown to be the curve $R=F$ and turns out to be the
 locus $z=1-2\lambda/3$.

The integral being over $z$, it is important to know which value of z along
 the outgoing null curve yields $r=0$, the lower limit of the integral. 
 This issue as it is shown further
 leads to the difference in the behaviour of $Q$ in the naked and covered cases. 

 Consider then the equation
\be
\label{r} r=e^{\left[-\int\frac {db}{b-R'(b)}\right]_{z}}
\ee
 which we examine for $r=0$. That will happen when the integral in the
 square bracket diverges positively. Two cases can be immediately seen to
 arise. 

\underline{Case(i)} $b-R'(b)=0$ has no real root.\\
The integrand therefore does not diverge anywhere and also remains positive
 (or entirely negative) all over the real line. It can be checked that
 $b-R'(b) > 0$ for any one real $b$ which would be sufficient to claim that
 the integrand is positive. Also, $b-R'(b)$ is bounded since $b$ is to
 be limited to the nonsingular region $z<1$ ($z=1$ is the singularity 
 curve itself). So, in order that the integral diverge, the range of 
integration should be infinite. We have chosen to limit the final point to the
 apparent horizon $z=1-2\lambda/3$ and hence the initial point must be 
$z=-\infty$.

 In fact a shorter intuitive argument is possible. It is known from the Tolman
 Bondi dust model that central singularity forms at ($t_{0}(0)$,$0$).
 If one assumes the Penrose diagram for the covered case (which is indeed
 what this case turns out to be), the null rays crossing the apparent
 horizon begin at the centre at $t<t_{0}$ which makes $z=-\infty$ there.

 \underline{Case(ii)} $b-R'(b)=0$ has at least one real root.

 In this case, the equation implies that $r=0$ at the value of $z$ for
 which the integral in the exponent diverges. The range of integration
 for equation (\ref{1/Qint}) would then be limited at the lower end by that value of $z$. This will
 be the root (in fact the one closest to $1-2\lambda/3$, the apparent horizon).

That this is so is seen as follows.     
Equation (\ref{r}) can then be re-cast using an expansion for the integrand in
 the exponent as outlined below.

Expanding R'(b) using the Taylor series about the root (called $z_{-}$)
 it can be shown that the leading order behaviour of r is as follows

 \be
r= (z-z_{-})^{\frac{1}{1-\alpha}} + O(z-z_{-})
\ee
 where 
\be
\alpha = \left(\frac{dR'(b)}{db}\right)_{b=z_{-}}
\ee

(It can be easily checked that $\alpha <0$)
   
 At $z=z_{-}$, therefore, $r$ vanishes.

Thus in conclusion of this analysis, we note that the lower limit of
 integral for $1/Q$ differs. It is $-\infty$ when $R'(b)-b=0 $ can never have
 a real solution and is the root (closest to apparent horizon) when a 
solution exists. 

This observation plays the key role in further analysis. Making note of this
 consider equation (\ref{1/Qint}). Analyzing the various factors
 in the integrand one finds that the integrand would diverge if $z=-1$
 ( $z \ne 1 $ since we are not on the singular boundary). 
 
 $z$ will take the value $-1$  in case i. In case ii,
 the following takes place. Consider $b-R'$ using equation (\ref{R'eqn}). It is easy
 to see that $ b-R' < 0$ for all $b<0$. So, the root $b_{-}$ cannot be
 negative. Hence it is certainly greater than $-1$. Thus $z$ cannot take the
 value $-1$ in case ii in the integral for $1/Q$.

Thus the integrand diverges as $1/(z+1)$ in case i and is finite in case ii.

Expanding the rest of the integrand factor in a Taylor series about $z=-1$,
 one can easily check that the integral diverges logarithmically in case i
 while staying finite in case ii.

Thus, $Q$ vanishes in case i and stays non zero (and finite) in case ii.

    Returning now to equation (\ref{xieqn}) ,  we can now see that {\bf $\xi$} behaves 
 in different ways in case i and case ii on the apparent horizon, in particular
 as  one approaches the point O on the Penrose diagrams shown (figures 3 and 4)
\footnote{Figure 4 is a diagram of a locally naked singularity. The self similar cloud which we examine here turns out to be globally naked. However, the
 structure near O is the same as any locally naked case and figure 4 can be used.}
. It can be checked that the factor $\sqrt{1+f}/R'$ tends to is a non-zero
 finite quantity on the apparent horizon.

  So, {\bf $\xi$} vanishes in case i and tends to  a non-zero quantity
  in  case ii. 

  Thinking of {\bf $\xi$} as `flux density' of the congruence, we may interprete
 that the congruence tends to cluster in case ii as against case i. 

  From previous analysis of naked singularities (self similar cases) using
  analysis for emergence of  geodesics (roots analysis), it can be checked
 that case i corresponds  to the covered case and case  ii  corresponds to
  the naked  singular  metric.

 \section{ Extension to the general dust case}

 In the general  dust case,  the equation (\ref{Qeqn}) yields no closed analytic
 solution which would have clearly been  useful. However, we note that
 we  are interested only  in the  behaviour of $Q$ in the limit
 of approach to point O on the apparent horizon.

To this end the following observation plays an important role. It is
 shown that given a dust solution, one can construct a modified dust
 solution (modified distribution) which in a  suitable limit approaches
 the given dust solution \cite{divpap}.  The key result that makes
 this construction useful is that it is proved that naked  modified
 distributions reproduce naked dust solutions given and  covered modified
 distributions reproduce covered ones. One can then work with the
 modified distribution for  the  given dust solution and take the limit
 which preserves naked or covered nature. We outline the construction
 in \cite{divpap} below

 a) Marginally bound case  ($f =0$)

   Imagine a shell of radius $r_c$ in  the given  Tolman Bondi dust
 model. Replace the interior of the shell by a  self similar dust metric,
 matching the first and second fundamental forms at the interface  $r=r_c$.
 It can be shown  that  this restricts the self similarity parameter 
 $\lambda$ which  appears in the mass function.  This specifies the
 self similar solution  completely. Now taking the limit  as $r_c$ tends to
 zero, one can  show \cite{divpap} that the  matching constraint does
 imply that the interior self similar solution stays naked in the limit
 if the original dust solution was naked and likewise in the covered case.

b) Non Marginally bound case ($f\ne 0$)

    The construction is similar in this  case except for  an additional  
 interface.  Two shells, $r_c1$ and $r_c2$ ( say  $r_c1 < r_c2$ )  are  now
 considered. To the interior of $r_c1$,  we replace by a self similar metric.
 Between  the two shells, we replace  with  a dust portion having  $f$ so
 behaved that it increases smoothly from zero at $r_c1$ to $f(r_c2)$ of the 
 original dust metric. The $F$ function for this extra portion of
 dust however  is  the same as that of the original dust metric.
 We match the first and second fundamental forms
 at each of the interfaces. As  before, this can be shown to  constrain 
 the interior self similar solution uniquely given $r_c1$ and the original
 dust solution. Again, the property of being naked or covered is preserved
 in the limit ( $r_c2 \rightarrow 0$) like the previous case  \cite{divpap}.

We now consider $Q$ in the modified distribution for  any  given dust
  solution. In  the self similar  part  of the latter, results of the
 previous section  apply. Since the congruence of  outgoing geodesics is
 smooth, so  is $Q$. This makes  $Q$ continuous across the interface/s
 in the modified distribution. Now imagine the given dust solution as
 the  limiting case of the modified distribution. In the limit of approach
  to point  O on the apparent horizon, one  has to evaluate $Q$ in the
 self similar part.  Because of continuity of $Q$, the same  behaviour
 will continue to hold  in the limit of the interface/s tending to zero  when 
  the original dust  solution is reproduced. Making use of the fact that
 the property  of being naked or covered is preserved in this limit, one
 concludes that the behaviour of $Q$  (and hence that of {\bf $\xi$})in the self similar naked  and covered
  cases continues to hold in the general dust scenario  as  well.

\section {Conformal transformation and Penrose diagram}

 The  issue about the metric being causally appropriate  
 is related to the behaviour of {\bf $\xi$} by the simple argument
 below. 

 The point O being the intersection of the apparent horizon and the singular boundary implies that any outgoing null congruence ought to have zero expansion
 in the limit of approach to O. So O ought not to have any outgoing geodesics 
emanating from itself. Hence {\bf $\xi$} ought to vanish if this causal property
 is correctly reflected in the metric. If the metric is not the correct one,
 then one performs a conformal transformation which diverges at O. This
 makes the {\bf $\xi$} vanish and yields the causally correct metric. 

 We  now show that the under a conformal transformation which diverges
 in the limit of O, {\bf $\xi$} which tends to a non-zero limit transforms
 to a vector field which  vanishes in the  limit. 

 Recall  that we defined {\bf $\xi$} for any geodesic  congruence using
 an affine parametrization. Under conformal  transformations,  affine parameters along null geodesics change ( unlike timelike geodesics which do  not
 remain geodesic curves, null  geodesics do stay so provided  the affine parameter changes appropriately). Infinitesimal parameter $ds$ transforms to
 $\Omega^{2}(x^{\mu})ds$ \cite{waldgr}, where $\Omega^{2}$ is a conformal transformation. Thus it is obvious that 
 $\xi^{\mu}=dx^{\mu}/ds$ if finite and  non-vanishing in the limit will
 vanish under $\Omega^{2}$ transformation provided the  latter diverges there.
  
 Thus we find that in the dust case, one requires a conformal
 transformation which diverges on the apparent horizon in the  limit of approach
 to the singularity in the naked case as  against the covered case where
  the usual metric given in equation (\ref{gendust}) is appropriate to describe the singularity
 structure.\footnote{ It is evident from  the definitions that the difference 
in the behaviour of 
 tangent vector is due to
 difference in the metrics, one being conformally related to the other.
 This raises the question of the tangent vector being an appropriate 
 characterization
 of the causal structure. It certainly is not appropriate at any event 
 within the spacetime but in the limit of approach of the boundary
 its behaviour indicates if a metric with the correct limiting causal
 structure has been employed in its calculation.}

 This justifies the difference in the structure of the singular boundary near O in figures 3 and 4.
 In the naked dust case,  when  one  uses the
  metric used in the previous sections (figure 2), it can be shown  that  several radial 
 null  geodesics
  appear to emerge from the central singularity with the same tangent
   vector \cite{simple}. Thus the limiting behaviour is not correct for the
 set of outgoing radial null geodesics to form a congruence. i.e. the condition that there exists only one curve of the set passing through any point is violated in the limit of approach to the central singularity. The causally 
approriate structure of the central singularity in the naked case is therefore
 not a point, but a one dimensional curve. The diverging conformal 
transformation does precisely that leading from figure 2 to figure 4. 
The central singularity being transformed to the one dimensional naked singular portion shown)
  Thus this naked point on the central singularity
 is depicted as a curve, to obtain the Penrose diagram. No such divergence
 is required, in the covered case while going from figure 1 to figure 3 as
 is also implied by the tangent vector vanishing at O in the usual Tolman
 Bondi metric of figure 1.

\section {Arbitrariness in the definition of {\bf$\xi$}}

As defined earlier, {\bf$\xi$} is a tangent vector, defined using an affine parameter,to a congruence of null geodesics the congruence being outgoing to 
 a marginal trapped surface in the apparent horizon. Thus, the {\bf$\xi$}
 field is not unique. The arbitrariness is due to two reasons. Firstly,
 the congruence is not uniquely defined. Secondly, the affine parameter
 used in the definition is arbitrary upto a multiplicative and additive
 constant. 

In spherical symmetry, the first difficulty is remedied by choosing
 a null congruence which is radial on the apparent horizon. This choice is not
 possible in general. It will be shown in the sequel to this paper 
 that vanishing (or nonvanishing) of the vector field at O is independent
 of the choice of the null congruence. 

The second difficulty brings in an
 arbitrariness of a scalar multiple of the vector field. This scalar is 
 constant along any null geodesic of the congruence. In the approach
 to the boundary point O, we claim that this will not affect the
 vanishing (or non vanishing) of the field. The argument is as follows.
 
  Consider a vector field   {\bf$\xi_{1}$} which does not vanish at any boundary point P.
  Let $ \phi(x)$ be the value of the scalar at any point $x$ on the manifold
 which multiplies this vector field (because of the multiplicative arbirtariness of the affine parameter). $ \phi$ is constant along any null
 geodesic of the congruence. We assume that P has a neighbourhood which 
 intersects the physical spacetime (spacetime without boundary) in a convex
 normal neighbourhood. So, {\bf$\xi_{1}$} not vanishing at P would imply
 the existence of a null geodesic beginning from P \footnote{ This is similar to Thm 8.2.1 of \cite{waldgr}}. Now, if $ \phi$ vanishes
 at P then it should vanish all along the portion of the null geodesic in the
 convex normal neighbourhood. This would make the multiplicative scaling
 of the affine parameter zero all along the null geodesic, which is impossible.
  So $ \phi$ cannot vanish in the approach to P. This completes the argument.

\section{Summary and Conclusion}

 The tangent vector field to an affinely parametrized null geodesic congruence
 is 
 examined for behaviour on the apparent horizon (defined as the boundary of
 the four dimensional trapped region as in the first section) in the approach to
 the singularity (point O) in the dust collapse model. There is a correlation
 with the property of nakedness with this behaviour. Demanding that the
 vector vanishes at the covered point O forces the divergence of the conformal
 transformation at O which leads to the Penrose diagram for the naked scenario.
 Since the vector vanishes in the covered case, there is no such divergence
 and hence the Penrose diagrams in the two cases differ.
        
 Thus, we have shown that the information about whether the singularity formed in collapse is naked is contained at the intersection of the boundary of the (four dimensional) trapped region and singular boundary in the spherical dust case.

The procedure of checking if an
 appropriate conformal transformation is necessary does not involve 
 checking
 for emergence of causal curves from the singularity. So far, the latter has been 
 the only method for checking if a singularity formed in collapse is
 naked. The work presented in this paper suggests an alternative method and demonstrates
 its validity in spherical dust collapse.


\begin{thebibliography}{99}
\bibitem{schoen} R. Schoen and S.T. Yau {\it Commun. Math. Phys}(1984){\bf 90}  575.
\bibitem{penrose} R. Penrose in {\it Black Holes and Relativistic Stars} ed. R.M. Wald (1990) University of Chicago Press.
\bibitem{rev1}  for recent reviews of the status of naked singularities in
classical general relativity, and for references to the literature, see for
instance, A.Krolak {\it Prog. Theor. Phys. Suppl.} {\bf 136} (1999) 45; R. Penrose in {\it Black Holes and Relativistic Stars } ed. R. M.
Wald (Chicago University Press, 1998); R. M. Wald, talk given at the April,1997 meeting in Washington, D.C., gr-qc 9710068  (see also R.M. Wald in {\it Black Holes and Relativistic Stars}) ;P. S.
Joshi, {\it Global Aspects in Gravitation and Cosmology} (Oxford, 1993); T. P. Singh, {\it Gravitational Collapse, Black holes and Naked Singularities} Proceedings of the Discussion Workshop on Black holes, Satellite Meeting of GR-15 ( Bangalore, 1997); C. J.
S. Clarke {\it Classical and Quantum Gravity} {\bf 10} (1993) 1375; P. S. Joshi
in {\em Singularities, Black Holes and Cosmic Censorship} ed. P. S. Joshi
(IUCAA, Pune, 1997); T. P. Singh in {\em Classical and
Quantum Aspects of Gravitation and Cosmology} ed. G. Date and B. R. Iyer
(Inst. of Math. Sc., Madras, 1996).
\bibitem{waldgr} R.M. Wald {\it General Relativity} (1984) University of Chicago  Press.
\bibitem{haw} S.W. Hawking and G.F.R. Ellis  {\it The Large  Scale Structure of
 Spacetime} (1973) Cambridge University Press.
\bibitem{gara} P.R. Garabedian {\it Partial Differential Equations} (1964) Wiley
\bibitem{rootspap} T.P. Singh  and P.S. Joshi {\it Class. Quant. Grav.} (1996)
 {\bf 13} 559 and  references  therein.
\bibitem{divpap} Sukratu Barve, T.P.Singh and Cenalo Vaz {\it Phys. Rev. D} (2000) {\bf 62}  084021
\bibitem{simple}Sukratu Barve, T.P.Singh, Cenalo Vaz and Louis Witten {\it Class. Quant. Grav.}  (1999) {\bf 16} 1727
\bibitem{jhingan}S.Jhingan, P.S.Joshi, T.P.Singh {\it Class. Quant. Grav.} (1996) {\bf 13} 3057
\end{thebibliography}
 \end{document}